%% file: merge.tex
\documentclass[twocolumn,showpacs,english,prx,preprintnumbers,amsmath,amssymb,floatfix]{revtex4}

\usepackage{subfiles}

\usepackage[T1]{fontenc}
\usepackage[latin9]{inputenc}
\usepackage{dcolumn}
\usepackage{bm}
\usepackage{graphicx}
\usepackage{color}
\usepackage{esint}
\usepackage{babel}
\usepackage{amsmath,amssymb,amsthm,mathrsfs,amsfonts,dsfont}
\usepackage{slashed}
\usepackage{enumerate}
\usepackage{simplewick}

\begin{document}

\subfile{letter.tex}

\clearpage

\subfile{SuppMat.tex}

\end{document}

%% file: letter.tex
\title{Effective field theory approach to many-body localization}

\author{Alexander Altland$^1$ and Tobias Micklitz$^2$}

\affiliation{ 
$^1$Institut f\"ur Theoretische Physik, Universit\"at zu K\"oln, Z\"ulpicher Str.~77, D-50937 K\"oln, Germany\\
$^2$Centro Brasileiro de Pesquisas F\'isicas, Rua Xavier Sigaud 150, 22290-180, Rio de Janeiro, Brazil}

\date{\today}

\begin{abstract}
We construct an analytic theory of many-body localization (MBL) in random spin  chains. The approach is based on a first quantized perspective in which MBL is
understood as a localization phenomenon on the high dimensional lattice defined by
the discrete Hilbert space of the clean system. We construct a field theory on that lattice and apply it to discuss the stability of a weak disorder (`Wigner-Dyson') and a strong disorder (`Poisson') phase. 
\end{abstract}


\maketitle

When shielded against external environments (`baths') interacting and disordered many
particle quantum systems may enter a state of `many-body localization' (MBL), a phase
distinguished by the absence of ergodicity and the vanishing of transport
coefficients. First observed in the low temperature regime of an interacting Fermi
system~\cite{BaskoAleinerAltshuler,Mirlin}, MBL is now recognized as a widespread
phenomenon shown by fermionic~\cite{fermions1},
bosonic~\cite{bosons1,bosons2,bosons3}, and spin
systems~\cite{spin1,spin2,spin3,spin4,spin5,spin6,spin7,Imbrie} over extended
parameter ranges; the first experimental observation of a many-body localized phase has been
reported recently~\cite{experiment}.

We owe much of our understanding of MBL to theoretical work that puts the focus on \emph{fixed realizations} of a disorder configuration. MBL
is then diagnosed from the analysis of phase space decay rates inspired by early work
on single particle localization on random lattices~\cite{Anderson}, the real space RG
approach to random spin systems~\cite{Dasgupta80,Fisher92,Igloi05}, or
phenomenological modeling~\cite{phenomenology}. A recent proof of MBL~\cite{Imbrie} 
proceeds in terms of a sequence of local unitary transformations whose generator is a functional of the disorder configuration. While these approaches provide
convincing evidence for the formation of localized regimes and phase transitions, they do
not provide us with \emph{effective} theories of the MBL phase, i.e. descriptions
similar in spirit to the powerful Ginzburg-Landau type field
theories~\cite{Wegner,EfetovBook} of single particle localized phases. Judging from
experience with single particle localization the construction of such theories will
be a powerful aid in the  identification of universality classes, the
description of critical phenomena, and that of observables. In this paper we
introduce the foundations of such a theory.

Our approach differs from previous work in two important respects. First, we discuss
our model system -- a random spin $1/2$ chain with local interactions -- from a
\emph{first quantized} perspective in which its Hamiltonian is considered as a matrix
in Hilbert space. This formulation  brings us in a
position to apply established methods of transport on random lattices to the problem.
Second averaging over disorder is performed at an early stage of the construction.
From that point on the system is described by an interplay of the \emph{clean} spin chain Hamiltonian  with effective field degrees of freedom introduced by the disorder
average. Our main task will be the identification of `soft modes', i.e.
field fluctuations of lowest action which describe the physics at large distance
scales. We will discuss how these fluctuations describe an Anderson localized and
delocalized phase depending on whether we are in a regime of strong or weak disorder,
respectively. However, the description of the transition between the two phases,
which we argue is \emph{not} in the Anderson universality class, is beyond the scope
of the present analysis.

\emph{Model and qualitative considerations:} We consider a system of $L$ spins
governed  by a Hamiltonian $\hat H \equiv \hat H_1+\hat H_2$, with random one-body
$\hat H_1$ and deterministic two-body $\hat H_2$. Here, $\hat H_1\equiv \sum_{l=1}^L
\epsilon_{l} \sigma_{z, l}$ describes a random magnetic field in $z$-direction with
Gaussian distributed field strengths $ \epsilon_l$ of variance $\gamma^2$. The
(non-random)  two-body interaction $\hat H_2\equiv \sum_l \hat V_l+\dots$ contains a
spin-exchange term $V_l=v \sigma^+_l\otimes \sigma_{l+1}^-+\mathrm{h.c.}$, where
$\sigma^\pm = \tfrac{1}{2}(\sigma_x\pm i \sigma_y)$ in the standard Pauli basis, $v$
sets the interaction strength, and the ellipses denote integrability breaking
contributions
-- such as rotational symmetry breaking magnetic fields or next nearest neighbor
   interactions -- not specified explicitly. The assumed non-integrability of the
\emph{clean} system will facilitate our later discussion of the
strongly interacting regimes.

Before turning to the field theoretical description of the system let us formulate a brief  synopsis. The key players of our analysis will be fluctuation amplitudes $Q_{nm}\sim \bar\psi_n\psi_m$ describing the phase coherent propagation of pairs of many-body wave function amplitudes $\psi_n$, where $n$ are spin-$z$ states $n=(n_1,\dots,n_L)$, $n_l\in
\{0,1\}$. In a single particle problem with independently distributed site diagonal 
disorder  (and likewise in the many-body `random energy model' considered in~Ref.~\cite{REM}) such
fluctuations  would be strictly confined in Hilbert space, $Q_n\equiv Q_{nn}$.
However the scarcity of independent disorder amplitudes in the MBL problem gives rise
to off-diagonal fluctuations subject to damping that grows continuously in $|n-m|$,
where $|n-m|=\tfrac{1}{2}\sum_l |n_l-m_l|$ is the Hamming distance between sites. These non-local fluctuations are responsible for the main differences between conventional and many body localization.

In
the absence of interactions, sites are strictly uncorrelated and the independent
fluctuations of $Q_{nn}$ describe a many-body Poisson phase with maximal
localization. Interactions play the role of a `hopping operator' in the lattice of
sites. If we treat the problem  ignoring the off-diagonal modes, it  reduces to a
conventional Anderson localization problem in a lattice whose coordination number
$\sim L$ is set by the number of states $m$ connected to a given $n$ by nearest
neighbor interactions. The extensive growth of the effective dimensionality of the
lattice then leads  to the incorrect prediction of
\emph{de}localization, no matter how weak the interaction. (This phenomenon has indeed been observed numerically in the random energy model~\cite{REM}.)

Below we will discuss how the inclusion of off-diagonal
fluctuations is the key to rectifying the picture  both in the weak and the strong
interaction phase. For weak interactions they describe an essentially important
parametric renormalization of hopping parameters, and a perturbative elimination of
`hopping operators' similar in spirit to that employed in Imbrie's proof of MBL in
Ising chains~\cite{Imbrie}. We will show how an analogous
weak rotation of basis in Hilbert space can be applied to perturbatively eliminate
the nearest neighbor hopping operators in the field theory describing the disorder
averaged system. This establishes the stability of the localized phase relative to
interactions. In the complementary limit of strong interactions  off-diagonal modes are required to correctly describe the translational invariance of the clean system and the formation of an ergodic Wigner-Dyson regime when weak disorder is turned on.

\noindent \emph{Field theory construction:} Localization properties of the system may be probed by products of advanced and retarded Green functions $G^\pm(n,m,\epsilon\pm \tfrac{\omega}{2})\equiv \langle n|(\epsilon^\pm
\pm\tfrac{\omega}{2}-\hat H)^{-1}|m  \rangle $  at weakly different energies $\epsilon\pm \omega/2$, respectively. Such products can be obtained from the generating functional
$
Z\equiv \int D(\bar\psi,\psi)\,\langle \exp(i\bar \psi(\hat \epsilon-\hat H)\psi)\rangle,
$
where $\psi=\{\psi^a_n\}$ is a $4\times 2^L$ dimensional supervector, and the four
component index $a=(\lambda,s)$ comprises an index $s=\pm$ discriminating between
retarded and advanced Green functions and between complex commuting
($\lambda=0\equiv \mathrm{b}$) and Grassmann components ($\lambda=1\equiv \mathrm{f}$), respectively.
The Green function energy arguments are contained in  $\hat \epsilon\equiv \epsilon+ \tfrac{1}{2}\omega^+\tau_3$ where $\tau_3$ is a Pauli matrix acting in advanced/retarded space. For simplicity we will focus on the band center throughout, $\epsilon=\omega=0$, in which case $\hat \epsilon= i \delta\tau_3$ merely contains an infinitesimal imaginary part. Green functions are obtained from $Z$ by differentiation w.r.t. suitably introduced sources, however, we suppress these for the sake of clarity throughout. Following standard procedures~\cite{EfetovBook} we average the functional over the Gaussian fluctuations of $\epsilon_l$ to generate a quartic term in $\psi$ which in a second step is decoupled by means of a supermatrix Hubbard Stratonovich field 
$\hat A=\hat A_{nm}^{ab}$ comprising commuting and anti-commuting elements.  After integration over the then Gaussian $\psi$-fields, 
the effective $A$-action reads 
$S[A]=\frac{1}{2 \gamma^2}\mathrm{str}(A_{nm}A_{mn})f_{nm}+
\mathrm{str}\ln\big(\hat \epsilon-\hat H_2+  A \big)$,
where the supertrace~\cite{EfetovBook} `str' extends over all indices not shown explicitly and the weight function $f_{nm}\equiv 1/(L-|n-m|)$. 

We proceed to subject the action to a stationary phase analysis 
and seek for solutions  of the equation $\delta S[A]/\delta A_{nm}^{ab}=0$, or
\begin{align}
\label{eq:AMeanField}
f_{nm}A_{nm}^{ab}=-\gamma^2\left(\frac{1}{\hat\epsilon -\hat H_2+A }\right)^{ab}_{nm}.
\end{align}
The structure of the equation tells us that the mean field configuration  plays the role of a `self energy' describing the influence of the disorder. Indeed, it is straightforward to verify that the equation is solved by the fully diagonal configuration   $A=i \kappa  \tau_3\otimes \openone$ where $\openone$ is the unit-matrix in Hilbert space, the value of the energy scale  $\kappa$ depends on the regime we are in and the sign of $\pm i \kappa$ is determined by the causality $\pm i \delta$ of the Green function. The physics of (de)localization is encoded in soft fluctuations around the diagonal configuration. Depending on the relative strength of disorder and interactions, these fluctuations are determined by the condition of approximate commutativity with the interaction operator $\hat H_2$, or the quadratic weight governed by the correlation function $f_{nm}$, respectively. We first discuss the  more involved latter regime, $\gamma\gg v$.

\noindent \emph{Strong disorder:} Referring to the supplementary material for details we note that for $\gamma\gg v $ the strength of the impurity self energy is set by  $\kappa\simeq \gamma\sqrt L$, i.e. the sum of $L$ random numbers $\pm \epsilon_{l}$. Due to the assumed weakness of the clean  Hamiltonian, $\hat H_2$, the mean field equations possess a large family of approximate solutions $A=i \gamma \sqrt{L} Q$, where $Q_{nm}\equiv Q_n\delta_{nm}$ are matrices site-diagonal in Hilbert space, $Q_n=T_n \tau_3 T_n^{-1}$, and $T_n=\{T_n^{ab}\}$ $4\times 4$-supermatrices describing fluctuations away from the diagonal $\tau_3$. Substitution of these configurations into the action shows that in  the non-interacting limit, $v=0$, the fluctuations $T_n$ fully cancel out. A straightforward expansion to  leading (quadratic) order in the interaction leads to the soft fluctuation action
\begin{align}
\label{eq:STdiag}
    S[Q]=
\frac{v^2}{2\gamma^2 L}
\sum_{nm}\mathrm{str}(Q_n Q_{m})\,X_{nm} 
\end{align}
where $X_{nm}=v^{-2}|\langle n|\hat V|m
\rangle|^2$ is a connectivity matrix assuming the value $1/0$ if two states $n, m$
are coupled/not coupled by exchange interactions. This action is equivalent to that
of an a-periodic Anderson lattice with sites, $n$, and bond connectivities $X_{nm}$.
For each $n$, we have $\mathcal{O}(L)$ non-vanishing elements $X_{nm}=1$ meaning that
the lattice has characteristic coordination number $Z=L$ and hopping strength
$\alpha\equiv v^2/\gamma^2 L$. Lattices of this type have an Anderson metal-insulator
transition at $\alpha_c$ determined by the equation $Z\sqrt{\alpha_c/2  \pi} \ln
(\alpha_c/2)  \sim (v/\gamma) L^{1/2}=1 $~\cite{EfetovBook}, i.e. the action~\eqref{eq:STdiag} predicts
Anderson \emph{delocalization} in the thermodynamic limit, $L\to \infty$, no matter
how weak the interaction. This result is in conflict with our understanding of MBL,
and it means that the restriction to diagonal fluctuations commutative with the
disorder weight must have been premature.

\begin{figure}
\centering
\includegraphics[width=6cm]{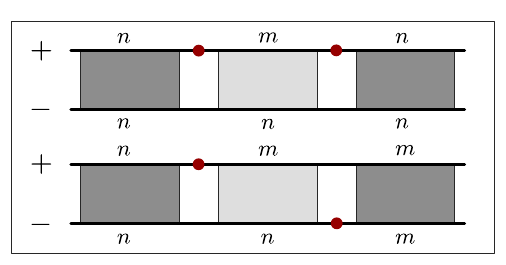}
\caption{\label{fig:ModeCoupling} (Color online) The coupling of modes, $T_d$, diagonal in Hilbert space (dark shaded) to weakly off-diagonal modes, $T_v$ (light shaded). Lines indicated by $+/-$ denote wave function amplitudes stationary at sites $n$ or $m$. Fluctuations of the respective modes represent quantum interference between these amplitudes. Further discussion, see text.}
\end{figure}

The key to resolving the situation lies in the observation that the model supports a large number of nearly soft modes, weakly non-commutative with both the interaction and the disorder weight. While a full integration over the effectively bi-local field of fluctuations $T_{nm}=\{T_{nm}^{ab}\}$ is  impossible the situation greatly simplifies if the interactions are treated perturbatively. Specifically, to lowest order in perturbation theory in an interaction channel $n\stackrel{X}\to m$ only fluctuations connecting to the sites $n,m$ are coupled (cf. Fig.~\ref{fig:ModeCoupling}, technically, all other fluctuations vanish by supersymmetry.)  We organize the corresponding fluctuation field as $Q\equiv T_s T_d T_v \tau_3 T_v^{-1} T_d^{-1}T_s^{-1}$, where the three fluctuation matrices $T_{s,d,v}$ play distinct physical roles: the center piece, $(T_d)_{lj}\equiv \delta_{lj}( T_{n}\delta_{jn}+T_{m}\delta_{jm})$ describes fluctuations local at the sites $n,m$ as considered before. The `vertex-flucutations' $T_{v}$ describe correlated fluctuations at the nearest neighbor sites $n,m$. These fluctuations may be represented as $T_v=\exp(W)$, where the fluctuation generator $W=\left( \begin{smallmatrix}
    &B\cr \tilde B
\end{smallmatrix} \right)^{\mathrm{ar}} $, $B=B_{nm}+B_{mn}$, $\tilde B=B^\dagger
\sigma_3^{\mathrm bf}$ (all other Hilbert space matrix elements vanishing) has an
off-diagonal block structure both in advanced/retarded, and $n,m$-space. In a manner
to be discussed momentarily, fluctuations of $T_v$ renormalize the second order
interaction vertex as indicated by the light shaded areas in
Fig.~\ref{fig:ModeCoupling}. Finally, $T_s=\exp(G)$ is a similarity transformation
generated by $G=\alpha \tfrac{v}{\gamma}(|n\rangle \langle m|+|m\rangle \langle n|)$,
where $\alpha$ is a tunable numerical parameter. This transformation plays a role
analogous to that employed in Imbrie's proof of MBL in Ising chains. It will be
employed to eliminate the effective interaction vertex by a Schrieffer-Wolff type
transformation \emph{after} the vertex fluctuations are integrated out.

We substitute the above configuration and expand in the coupling between sites $n,m$
to second order in $v/\gamma$, i.e. in $\hat H_2$, and $\hat G$. To zeroth order in
the interaction, both $T_s=1$, and $T_d$ cancels out in the disorder vertex. However,
the presence of the function $f$ implies that the vertex fluctuations get weighed
with an action $S_0[Q_v]=\tfrac{1}{4L}\mathrm{str}(Q_v\tau_3Q_v\tau_3)$, where
$Q_v=T_v\tau_3 T_v^{-1}$. In a similar manner (cf. the supplementary material) we obtain the coupling to the
interaction vertex as (indices suppressed for clarity)
$S[Q,X]=\tfrac{iv}{\gamma\sqrt L}\mathrm{str}(T_d^{-1}XT_d
Q_v)+\tfrac{v^2}{2\gamma^2 L} \mathrm{str}(T_d^{-1}XT_d Q_v)^2$ and to the generator
of the similarity transformation $S[Q,G]=\tfrac{1}{L}\mathrm{str}(T_d^{-1}GT_d
\tau_3Q_v )^2$.
The vanishingly small weight
$L^{-1}$ multiplying the vertex fluctuations means that the integral over the
non-linear super-manifold spanned by the matrix $Q_v$ has to be done rigorously.
Referring for details to the supplementary material, we here merely state  the
remarkably simple result of the integration,
$S_{\mathrm{eff}}[Q,G]
=
\tfrac{\sqrt{\pi} v^2}{\sqrt L
\gamma^2}\mathrm{str}(Q_n Q_m)X_{nm}
-
\tfrac{\sqrt{\pi}}{\sqrt{L}}\mathrm{str}(QG QG)$, 
i.e. the action~\eqref{eq:STdiag} multiplied
by a factor $\sqrt{L}$ plus a term coupling to the generator of the similarity
transform. This effective action affords a straightforward physical interpretation:
the dimensionless coupling constants, $c$, of site-to-site hopping terms in the field
theories of disordered systems are products, $c=\rho_0 \Gamma $ of the local (i.e. on-site) density
of states of the disordered non-interacting system, $\rho_0$, and a characteristic hopping rate $\Gamma$. Presently, $\rho_0\sim
1/\sqrt{L}\gamma$ is given by the ratio of one over the characteristic
band-width. The hopping rate  due to interactions equals $v^2/\gamma$, where
$\gamma\sim |\epsilon_n-\epsilon_m|$ is the characteristic energy difference between
two levels of Hamming distance $\mathcal{O}(1)$. Technically, the full integration
over the vertex fluctuations was required to resolve this small energy difference
with accuracy.

We may now chose a perturbative basis rotation $\mathds{1}+G\equiv
\mathds{1}+\tfrac{v}{\gamma} X_{nm}$ to effect a mutual cancellation of the two
terms in the action. This should be compared to the construction of Imbrie in which a
similarity transformation generated by  $v^2 \tfrac{X_{nm}}{\epsilon_n - \epsilon_m}$
was applied to eliminate a spin-flip operator in the \emph{unaveraged} model of an
Ising chain~\cite{Imbrie}. In effect, our approach achieves a construction of similar nature
within the framework of an effective averaged theory. A perturbative extension to
higher orders in $v/\gamma$ may be applied~\cite{future} to eliminate site hopping over larger
distances at higher order in the $X$-expansion. Evidently, this scheme breaks down at
values $v>\gamma$ when a transition outside the standard Anderson universality class
is expected to take place. (Within the field theoretical framework, an Anderson
transition would be described by local hopping operators as in~\eqref{eq:STdiag}
which the above construction shows are not present in this form.) Before
discussing the situation in the complementary limit of strong interactions, it is
worth comparing to the random energy model which has been introduced as a
phenomenological model of many-body Hilbert space localization in Ref.~\cite{REM}. In that
model, the on-site energies $\epsilon_n$ are chosen as $2^{L}$ independent random
variables and the ensuing field theory is described by strictly local fluctuation
modes, $T_n$. The absence of nonlocal modes means that Eq.~\eqref{eq:STdiag}
describes the theory at weak interactions and puts it in the Anderson universality
class. As argued above, and in line with numerical observation this implies
delocalized behavior in the large system size limit at arbitrarily weak interactions.
The difference to the genuine MBL universality class with its instability at
$v/\gamma=\mathcal{O}(1)$ is related to the scarcity of independent disorder
amplitudes in the latter and the emergence of non-local quantum interference modes.

\noindent \emph{Strong interaction:} We next explore what happens in the complementary 
limit $v/\gamma\gg 1$ of the strongly interacting system.  
In this case, a spectral decomposition of the r.h.s. of
Eq.~\eqref{eq:AMeanField} in the eigenfunctions of $\hat H_2$ shows that the impurity self energy 
$\kappa=\pi \gamma^2 \rho$ multiplying the solutions $A$ of Eq.~\eqref{eq:AMeanField} is given 
by a golden rule product of the on-site band center density of states of the clean interacting system, $\rho \sim 1/Lv$, 
and the scattering probability $\sim \gamma^2$ (see supplementary material). 
Low action fluctuations $T$ around the diagonal solution must be commutative with the interaction Hamiltonian $[T,\hat H_2]=0$. For a generic (non-integrable) $\hat H_2$ a set of modes satisfying this condition can be constructed by switching to a basis~\cite{Bogomolny} $n\equiv (\bar n,s)$ in which each site $n\equiv \hat T^s\bar n$ is represented as an $s$-fold translation of a site $\bar n$ of a $2^L/L$-dimensional unit cell,  $n_l=(\hat T^s \bar n)_l\equiv (\bar n)_{l-s}$.  It is then straightforward to check that modes $T_{n,m}\equiv T_{(r,\bar n),(s,\bar m)}\equiv T_{r-s}\delta_{\bar n,\bar m}$ depending only on the translation sector commute with the translationally invariant interaction Hamiltonian. The substitution of $A=i \kappa T\tau_3 T^{-1}$ into the disorder weight $S_{\rm dis}[A]=(1/2 \gamma^2)\sum_{nm} \mathrm{str}(A_{nm}A_{mn})f_{nm}$ leads to a term coupling these modes, $S_{\mathrm{dis}}[Q]=\mathrm{const.}\times  \bar\rho \tfrac{\gamma^2}{v}\sum_r \mathrm{str}(Q_r Q_{-r})f(r)$, where the sum is over translation sectors, and the effective  weight function $f(r)=\tfrac{L}{2^L}\sum_{\bar n}f_{(\bar n,r),(\bar n,0)}$ measures the characteristic Hemming distance between translated states. The coupling constant can be interpreted as the product of the on-site density of states on the lattice of translational modes,  $\bar\rho=\rho\, 2^L/L$, and the golden rule scattering rate scattering rate between clean eigenstates of the system. Due to the large  density of states $\bar\rho\sim2^L$ disorder strengths exponentially small in system size suffice to effect a freezing of the translation modes to a single effective zero mode $T_{nm}\equiv T_0\delta_{nm}=T_0 \delta_{\bar n,\bar m}\delta_{r, s}$ fully diagonal in Hilbert space. This mode has vanishing action and describes fully ergodic behavior at large interactions $v>\gamma$.

\noindent \emph{Phenomenological consequences:} The differences between the regimes
discussed above show in the behavior of system observables, for example in its many
body spectral correlations. In the clean limit, $\gamma=0$, the $L$ translation
sectors of fluctuations $T_r$ describe an equal number of Hilbert space sectors
irreducibly transforming under the translationally invariant Bloch Hamiltonian. Each
sector individually shows Wigner-Dyson correlations (as described by the fluctuations
of the corresponding translation mode), however, the lack of statistical correlations
between them implies that the many-body spectrum of the nearly clean system does not
show level repulsion. For disorder strong enough to couple the translation sectors,
and up to values $ \gamma\lesssim v$ a Wigner-Dyson regime described by the
fluctuations of a fully ergodic mode $T_0$ ensues. The transition region
$\gamma\simeq v$ is beyond the control of the present theory. However, for strong
disorder $\gamma\gtrsim v$, the perturbative decoupling of the fluctuations $T_d$,
which describes the statistical independence of fluctuations in the locally
diagonalized Hilbert space basis, implies Poissonian statistics. Both, the
Wigner-Dyson and the Poissonian forms of the spectral statistics can be explicitly
derived by integration over the corresponding field modes using the techniques
described in Ref.~\cite{EfetovBook}.

\noindent \emph{Discussion:} We have derived an effective field theory describing the
physics of $XXZ$-chains in the presence of local disorder. Formulated in a first
quantized language the theory effectively  describes quantum  transport on a
translationaly non-invariant lattice subject to site-diagonal
disorder. Striking differences to standard single particle Anderson localization lie
in the extensively high dimensionality $\sim L$ of the lattice, and in the scarcity
of only $\mathcal{O}(L)$ independent disorder parameters on a lattice with an
exponentially large number $2^L$ of sites. Where the high configuration number favors
delocalization, the statistical dependence of the site disorder leads to the
formation of off-diagonal fluctuation modes describing statistical correlations
between different sites. These fluctuations support localization via an effective decoupling of sites which we described to leading order in perturbation theory in $v/\gamma$, and which we argued
puts the system outside the Anderson class. As a sanity check we compared to the
phenomenological random energy model which lacks this mechanism and in the
consequence shows a delocalization instability at arbitrarily weak interactions. It is also tempting to compare the fluctuations of the theory in the effectively decoupled frame to those of the phenomenological theory of $l$-bits~\cite{phenomenology}, however further work is required to substantiate this picture. 

\noindent \emph{Acknowledgments:} A.A. acknowledges discussions with  J. Chalker, C. Laumann, V. Oganesyan, M. M\"uller, and A. Scardiccio, and the hospitality of the workshop AWGMBL, Cambridge 2016. Work supported by the program ``Science Without Borders" of CNPq (Brazil) and CRC 183 of the Deutsche Forschungsgemeinschaft. 
T. M. acknowledges financial support by Brazilian agencies CNPq and FAPERJ.

%% file: SuppMat.tex
\title{Supplementary Material to ``Effective field theory approach to many-body localization"}

\author{Alexander Altland$^1$ and Tobias Micklitz$^2$}

\affiliation{ 
$^1$Institut f\"ur Theoretische Physik, Universit\"at zu K\"oln, Z\"ulpicher Str.~77, D-50937 K\"oln, Germany
\\
$^2$Centro Brasileiro de Pesquisas F\'isicas, Rua Xavier Sigaud 150, 22290-180, Rio de Janeiro, Brazil}

\date{\today}

\begin{abstract}

In this supplementary material we discuss solutions to the mean field equation in the limits of weak and strong disorder, 
and provide details on the renormalization of the second order interaction vertex.
\end{abstract}


\maketitle

\section{Mean field equation}

We here discuss solutions to the mean field equation (1) of the main text. We start from a matrix-diagonal ansatz $A_{nm}=i\kappa\tau_3 \otimes \mathds{1}^{\mathrm bf} \delta_{nm}$, $\kappa>0$, homogeneous in boson-fermion and Hilbert-space, and where the sign change described by $\tau_3=\tau_3^{\mathrm{ar}}$ is dictated by the causality of the Green function. At the band center the equation then assumes the form
\begin{align}
\label{appMF}
A=-\gamma^2L \left(\frac{1}{i\delta \tau_3-\hat H_2+A }\right)_{nn},
\end{align}
where  we used that $f_{nn}=1/L$. {\it Strong disorder: } In the strong disorder limit one may neglect the interaction, $\hat H_2=0$, to obtain a quadratic equation which is solved by $\kappa=\gamma \sqrt{L}$. 
{\it Strong interaction:} In the strongly interacting limit, we may expand the equation in
 eigenfunctions $\{|\alpha\rangle\}$ of $\hat H_2$,
\begin{align}
\label{appMFsInt}
A
&=-\gamma^2L \sum_\alpha {|\langle n|\alpha\rangle|^2 \over i\delta\tau_3 -E_\alpha + A}. 
\end{align}
Ignoring a real energy shift contribution to $A$ (which may be absorbed by a redefinition of the energy argument) we then find
\begin{align}
\label{eq:SpStrongI}
\kappa
&\simeq
\gamma^2\, \mathrm{Im}\int dE \frac{ \rho(E)}{E - i\kappa}
\simeq \pi \gamma^2\rho,
\end{align}
where $\rho\sim 1/Lv$ is the on-site density of states of the clean interacting system,
and we made the self consistent assumption that the disorder generated smearing of states, $\sim \kappa$, 
is weak enough that $\rho(E)$ remains approximately constant for energies at the band center.

\section{Renormalization of the interaction-vertex}

We present details on the derivation of the effective second order interaction-vertex by integration over the vertex mode.
Starting out from the action 
$S=S_w+S^1_X+S^2_X$, where $S_w
=
-{1\over 2L}
 \sum_{nn'} {\rm str}( Q_{nn'} f_{nn'}  Q_{n'n} )$ is the disorder weight and $S_X^1
=
{i v \over \gamma \sqrt{L}}\,
{\rm str}(X  Q)$ and $
S_X^2
=
{v^2 \over 2\gamma^2 L}\,
{\rm str}(X  Q X  Q)$, are the expansion of the tr ln to first and second order in the interaction, respectively, the goal is to integrate out fluctuations coupling  nearest-neighbor sites  
to arrive at an effective interaction between Hilbert-space diagonal modes.

To this end we fix a pair $n,m$ of nearest neighbors and 
parametrize fluctuations as discussed in the main text,
$Q= T_s T_d Q_v T^{-1}_d T_s^{-1}$,
where 
$Q_v=e^W\tau_3 e^{-W}$ with 
$W=(\begin{smallmatrix}
& B_{nm} \\
B_{mn}^\dagger\sigma_3^{\rm bf} & 
\end{smallmatrix})+(n\leftrightarrow m)$. The two contributions to the generator $W$ commute which means that the matrix $Q_v$ splits into two independent matrices with block elements   
\begin{align}
\label{appEq1}
(Q_v^{++})_{nn}, \quad
(Q_v^{+-})_{nm}, \quad
(Q_v^{-+})_{mn}, \quad
(Q_v^{--})_{mm},
\end{align}
and a second set of blocks with $n\leftrightarrow m$, respectively. The two $(n\leftrightarrow m)$ interchanged matrix sectors are independently generated and can be integrated separately. We focus on the first, and add the $(n\leftrightarrow m)$ contribution at the end of the calculation. 
A straightforward expansion of the action to second order in  $G\sim X$ then leads to 
\begin{align}
S^0_w
&=
 {1\over 4L}
  {\rm str}(Q_v\tau_3 Q_v\tau_3 )
\\
S^2_w
&=
-{1\over L}
 {\rm str}(\,
 2 \Phi_G P^- Q_v P^- \Phi_G P^+ Q_v P^+
 \nonumber \\
 & \qquad \qquad 
 -
  \Phi_G P^- Q_v P^+ \Phi_G P^- Q_v P^+
 \nonumber \\
 & \qquad \qquad
 -
  \Phi_G P^+ Q_v P^- \Phi_G P^+ Q_v P^-
  )
\\
S_X^1
&=
{i v \over \gamma \sqrt{L}}\,
{\rm str}(\Phi_X P^- Q_v P^+ + \Phi_X P^+ Q_v P^- )
\\
S_X^2
&=
{v^2 \over 2\gamma^2 L}\,
{\rm str}(
 2 \Phi_X P^- Q_v P^- \Phi_X P^+ Q_v P^+
 \nonumber \\
 & \qquad \qquad 
 +
  \Phi_X P^- Q_v P^+ \Phi_X P^- Q_v P^+
 \nonumber \\
 & \qquad \qquad
 +
  \Phi_X P^+ Q_v P^- \Phi_X P^+ Q_v P^-
),
\end{align}
where  
 $P^\pm={1\over 2}\left(\openone \pm \tau_3\right)$ 
 are projection operators in advanced-retarded space,
 and we introduced $\Phi_O=T^{-1}OT$. We note that $\Phi_{X,G}$ have the same $(+n,-m)$ Hilbert space structure as $Q_v$ in Eq.~\eqref{appEq1}, suppressed for notational simplicity.  

The effective interaction vertex between Hilbert-space diagonal modes at sites $n,m$,
is obtained by perturbative expansion in the interaction,
\begin{align}
\label{appEq2}
S_{\rm eff}
&=
\langle
S_w^2 + S_X^2  - {1\over 2} S_X^1 S_X^1
\rangle_{Q_v},
\end{align} 
where $\langle...\rangle_{Q_v}\equiv \int dQ_v\,  e^{-S^0_w}$. 
We are left with the task of doing  the integrals in \eqref{appEq2}. 
Following Efetov~\cite{EfetovBook} we parametrize $Q_v=U_2U_1Q_0U_1^{-1}U_2^{-1}$, where ($i=1,2$)
\begin{align}
Q_0
&=
\begin{pmatrix}
\cos\hat\theta & i \sin\hat\theta
\\
-i \sin\hat\theta & -\cos\hat\theta 
\end{pmatrix}_{\rm ra},
\quad
U_i
=
\begin{pmatrix}
u_i^+ &
\\
& u_i^-
\end{pmatrix}_{\rm ra},
\end{align}
and the supermatrices 
$\hat\theta=\rm{diag}(i\theta_{\rm b},\theta_{\rm f})_{\rm bf}$, 
$u^\pm_1=e^{-2\hat\eta^\pm}$ with
$\hat \eta^\pm
=
(\begin{smallmatrix}
 & \bar \eta^\pm
\\
- \eta^\pm & 
\end{smallmatrix})$,
$u^+_2
=
(\begin{smallmatrix}
e^{i\phi} &
\\
& e^{i\chi}
\end{smallmatrix})$,
and
$u^-_2
=
\openone$. Here $0\leq\theta_{\rm f}<\pi$ and $0\leq\theta_{\rm b}$ parametrize the compact 
fermionic and non-compact bosonic sectors, respectively, 
$0\leq \phi,\chi<2\pi$, and $\bar\eta^\pm,\eta^\pm$
are independent Grassmann variables.
The non-interacting action reads 
$S_w^0={1\over L}(\cosh^2\theta_{\rm b}-\cos^2\theta_{\rm f})$
and we notice that the main contribution to the integrals results from the large integration volume of
the non-compact bosonic angle, $\theta_{\rm b}\lesssim \ln L$.
We may thus approximate
$Q_0
=
(\begin{smallmatrix}
\cosh\theta_{\rm b} & -\sinh\theta_{\rm b}
\\
\sinh\theta_{\rm b} & -\cosh\theta_{\rm b} 
\end{smallmatrix})
\otimes P^{\rm b}$, 
where $P^{\rm b}={1\over 2}(1 + \sigma_3^{\rm bf})$ is a projector onto the bosonic sector. 
Using the measure of the above polar representation~\cite{EfetovBook}
$dQ 
= {1\over 2^6 \pi^2} 
{\sinh\theta_{\rm b} \sin\theta_{\rm f} \over (\cosh\theta_{\rm b}-\cos\theta_{\rm f})^2}
d\phi\, d\chi\, d\theta_{\rm b}\, d\theta_{\rm f}\,  d\bar\eta^+  d\eta^+ d\bar\eta^- \, d\eta^-$, and integrating over commuting and anti-commuting variables we find 
\begin{align}
\langle 
S^2_w
\rangle_{Q_v}
&=
2\sqrt{\pi\over L}
{\rm str}( \Phi_G P^- \Phi_G P^+ ),
\\
\langle 
S^2_X
\rangle_{Q_v}
&=
-\sqrt{\pi\over L} {v^2\over\gamma^2} 
{\rm str}( \Phi_X P^- \Phi_X P^+ ),
\\
\langle 
S^1_X
S^1_X
\rangle_{Q_v}
&=
\sqrt{\pi\over L} {2v^2\over\gamma^2}
{\rm str}( \Phi_X P^- \Phi_X P^+ ).
\end{align}
This can be rewritten as 
\begin{align}
\label{appEq3}
S_{\rm eff}
&=
{\sqrt{\pi} \over 2 \sqrt{L}} \left( 
{v^2\over \gamma^2}{\rm str}(Q_m Q_n )X_{nm}  
-
{\rm str}(G_{nm} Q_m G_{mn} Q_n ) 
\right),
\end{align}
where $Q_n=T_n \tau_3 T^{-1}_n$ are the Hilbert-space diagonal modes, 
and we re-introduced Hilbert-space indices. 
Adding the second contribution from
generators with $n\leftrightarrow m$ exchanged 
increases the action \eqref{appEq3} by a factor two.